\documentclass{article}
\usepackage{authblk}

\usepackage[utf8]{inputenc}
\usepackage{graphicx}
\usepackage{float}
\usepackage{cite}
\usepackage{url}
\usepackage{multirow}
\usepackage{indentfirst}
\usepackage[a4paper, left=1in, right=1in, top=1in, bottom=1in]{geometry}

\title{Resource-Aware Layered Intrusion Detection Allocation Model}

\author{Ioan P\u{a}durean}
\author{Béla Genge}
\author{Roland Bolboac\u{a}}
\affil{Department of Electrical Engineering and Information Technology\\
George Emil Palade University of Medicine, Pharmacy, Science, and Technology of Targu Mures, Romania.\\
\texttt{padurean.ioan.25@stud.umfst.ro}, \texttt{\{bela.genge, roland.bolboaca\}@umfst.ro}}

\begin{document}

\date{\today}
\maketitle
\begin{abstract}
This paper proposes a resource-aware allocation model for layered intrusion detection in heterogeneous
networks. Monitoring traffic at higher protocol layers improves the ability to detect sophisticated attacks,
but it also increases computational and storage costs. The problem is formulated as an integer linear program
that assigns a single monitoring depth, ranging from Ethernet to the application layer, to each device, while
accounting for device importance, attack probability, layer-dependent detection rates, and per-layer monitoring
costs. The model further enforces a global resource budget, a minimum monitoring level for critical devices,
and maximum-feasibility limits for constrained devices such as simple IoT sensors. The formulation is solved
with the SCIP optimization framework on a small heterogeneous network of six devices, and the resulting
allocation illustrates how the model concentrates monitoring effort on important and high-risk devices while
respecting feasibility and budget constraints.
\end{abstract}

\section{Introduction}
Intrusion detection in modern networks is no longer limited to a single observation point or a single protocol layer.
Ethernet-level inspection can reveal attacks such as MAC spoofing or ARP poisoning
\cite{majumder2025arp,hnamte2024enhancing}, while deeper inspection at the IP, transport, and application layers
can identify increasingly complex threats such as distributed denial-of-service campaigns, session hijacking, SQL
injection, or brute-force attacks \cite{ennaji2025adversarial}. The main difficulty is that deeper monitoring
provides stronger visibility at a significantly higher resource cost \cite{tekin2023energy}.

In practical deployments, networked systems contain devices with different operational roles, risk profiles,
and technical limitations. A database server, for example, is both more valuable and more exposed than a simple
IoT sensor, while some constrained devices do not produce application-layer traffic that would justify deep inspection
\cite{isong2024insights}. This motivates the need for an allocation strategy that selects the appropriate monitoring
level for each device while respecting global resource limits \cite{liu2023delay}.

This article formulates the problem as an optimization problem for resource-aware layered intrusion detection. The model
captures device importance, attack probability, monitoring cost, and detection efficiency, and it includes additional
constraints for critical devices and devices with limited monitoring feasibility.

\section{Related Work}
Dao et al.~\cite{dao2023optimal} address the problem of assigning network intrusion detection (NID) tasks
in multi-level IoT systems where edge-fog computation resources and network traffic vary dynamically. The
authors formulate the assignment problem as an integer linear program whose objective is to minimize NID
latency while satisfying detection accuracy requirements and computation resource constraints, and they
additionally propose three heuristic strategies to obtain near-optimal solutions: a shortest-detection-based
assignment; a nearest-neighbor-based assignment; and a genetic algorithm. Their evaluation on two real-world
IoT attack datasets reports the effectiveness of the proposed algorithms with respect to detection accuracy
and latency. The present work shares the perspective of treating intrusion detection deployment as a
constrained resource-allocation problem, but focuses on selecting a monitoring depth across protocol layers
rather than placing detection tasks across edge-fog nodes.

Yang and Shami~\cite{yang2025toward} propose a multi-objective AutoML-based intrusion detection system
designed to enhance both performance and autonomy in cybersecurity operations. Their framework leverages
automated machine learning techniques to simultaneously optimize multiple criteria, such as detection
accuracy, computational cost, and model complexity, reducing the need for manual feature engineering,
model selection, and hyperparameter tuning and aligning with the broader movement toward self-adaptive
and intelligent security systems. The framework integrates two main components: an automated feature
selection method (OIP-AutoFS) that reduces dimensionality by selecting the most informative features, and
a joint optimization process (OPCE-CASH) that simultaneously chooses the best learning algorithm together
with its hyperparameters. By optimizing several objectives instead of accuracy alone, the system is
intended to perform effectively in resource-constrained environments such as IoT and edge networks, and
the authors report that experiments on benchmark datasets show improvements over existing intrusion
detection methods in both performance and efficiency. This direction is complementary to ours: they
emphasize the automated construction and tuning of machine-learning-based detectors under multiple
objectives, whereas the model proposed here is concerned with the upstream allocation decision of how
deeply each device should be inspected given limited monitoring resources.

Ibrahim et al.~\cite{ibrahim2012adaptive} propose an adaptive, layered intrusion detection system (IDS)
that combines machine learning techniques with gain-ratio-based feature selection to improve detection
accuracy and efficiency. The system is structured in two stages: an initial layer that distinguishes
between normal and attack traffic, followed by multiple specialized layers that classify specific types
of attacks, such as DoS, Probe, U2R, and R2L. By applying gain ratio to select the most relevant features
for each layer, the model reduces dimensionality and computational overhead while enhancing classification
performance. The authors evaluate several machine learning algorithms within this framework and report
that the layered approach improves detection rates, particularly for difficult attack categories, while
maintaining adaptability and scalability compared to traditional single-stage IDS models. This staged
classification architecture conceptually motivates the layered view adopted in the present paper. In
contrast to their detector-side architecture, the contribution presented here addresses the network-side
decision of which layer of inspection to enable for each device under explicit cost and feasibility
constraints.

\section{Methodology}

This section defines the linear optimization problem for resource-aware layered intrusion detection allocation.
The problem aims to determine the optimal monitoring depth for each device in a heterogeneous network, balancing
detection efficiency against resource constraints, while ensuring that constraints pertaining to critical devices
and monitoring feasibility are satisfied.

\subsection{Problem Description}
In cybersecurity, monitoring traffic at higher layers, such as the IP, transport, and application layers, enables the
detection of more sophisticated attacks, but it also leads to higher resource consumption.
Therefore, the problem is to determine the monitoring level for each device in the network under the following
conditions, with the goal of optimizing resource usage and maximizing detection efficiency:
\begin{itemize}
    \item A set of devices connected to a network;
    \item Limited monitoring resources;
    \item Different monitoring costs for different layers;
    \item Different levels of importance and attack probability for each device.
\end{itemize}

The considered monitoring layers are the following:
\begin{itemize}
    \item Layer 1: Ethernet-level monitoring (e.g. MAC spoofing detection, ARP poisoning);
    \item Layer 2: IP-level monitoring (e.g. IP spoofing, ICMP flood, Ping of Death);
    \item Layer 3: transport-layer monitoring (e.g. DDoS attacks, port scans, TCP session hijacking);
    \item Layer 4: application-layer monitoring (e.g. SQL injection, cross-site scripting, brute-force attacks).
\end{itemize}

\subsection{Linear Optimization Problem for Layered Intrusion Detection Allocation}
Let $D$ denote the set of devices in the network.

\subsubsection{Sets}
\begin{itemize}
    \item $i \in \{1, 2, \ldots, n\}$: index of device $i$ in the network;
    \item $l \in \{1, 2, 3, 4\}$: index of the monitoring layer.
\end{itemize}

\subsubsection{Decision Variables}
\begin{itemize}
    \item $y_{i,l} \in \{0, 1\}$: binary variable indicating whether device $i$ is monitored up to layer $l$ (1 if yes, 0 otherwise).
\end{itemize}

The interpretation of the layer selection is:
\begin{itemize}
    \item $l=1$: Ethernet-level monitoring;
    \item $l=2$: Ethernet + IP monitoring;
    \item $l=3$: Ethernet + IP + transport monitoring;
    \item $l=4$: Ethernet + IP + transport + application monitoring.
\end{itemize}

\subsubsection{Parameters}
\begin{itemize}
    \item $w_i$: importance of device $i$. For example, a router is more important than a laptop, a laptop is more important than an IoT sensor, and a database server is more important than a router.
    \begin{itemize}
        \item Database server: $w_i = 10$;
        \item Router: $w_i = 8$;
        \item Matter door lock: $w_i = 6$;
        \item Laptop: $w_i = 4$;
        \item IoT sensor: $w_i = 2$.
    \end{itemize}
    \item $p_i$: probability of attack against device $i$. For example, Internet-exposed servers have a higher attack probability than internal devices.
    \begin{itemize}
        \item Database servers: $p_i = 0.6$;
        \item Router: $p_i = 0.4$;
        \item Matter door lock: $p_i = 0.8$;
        \item Laptop: $p_i = 0.2$;
        \item IoT sensor: $p_i = 0.3$.
    \end{itemize}
    \item $d_l$: detection rate at layer $l$, with $d_l < d_{l+1}$, meaning that higher layers detect a larger fraction of attacks.
    \begin{itemize}
        \item Layer 1: $d_1 = 0.2$ (20\% of attacks detected);
        \item Layer 2: $d_2 = 0.5$ (50\% of attacks detected);
        \item Layer 3: $d_3 = 0.8$ (80\% of attacks detected);
        \item Layer 4: $d_4 = 0.95$ (95\% of attacks detected).
    \end{itemize}
    \item $R$: total available resources for monitoring;
    \item $c_l$: monitoring cost for layer $l$;
    \item $C$: set of critical devices that must have at least a minimum monitoring level $\alpha \geq 2$.
    Examples include database servers, application servers, and the main router, which should be monitored at least up to layer 2;
    \item $F$: set of devices for which higher-layer monitoring is not feasible, with maximum admissible level $\beta \leq 3$.
    For example, simple IoT sensors may not be monitored at layer 4 because they do not generate application-layer traffic.
\end{itemize}

\subsubsection{Objective Function}
The objective is to maximize detection efficiency, modeled as:
\[
\max \sum_{i=1}^{n} \sum_{l=1}^{4} w_i p_i d_l y_{i,l}
\]

This formulation assigns a higher contribution to devices that are more important, more likely to be attacked, and monitored at deeper protocol layers.

\subsubsection{Constraints}
\begin{enumerate}
    \item The total monitoring resources must not exceed the available capacity:
    \begin{equation}
    \sum_{i=1}^{n} \sum_{l=1}^{4} c_l y_{i,l} \leq R
    \end{equation}

    \item Each device is assigned exactly one monitoring depth. A higher layer already implies visibility over lower-layer traffic, so selecting multiple layers for the same device is unnecessary:
    \begin{equation}
    \sum_{l=1}^{4} y_{i,l} = 1, \quad \forall i \in D
    \end{equation}

    \item Critical devices must be monitored at least up to level $\alpha$:
    \begin{equation}
    \sum_{l=\alpha}^{4} y_{i,l} = 1, \quad \forall i \in C
    \end{equation}

    \item Monitoring feasibility is limited by the maximum level $\beta_i$ for constrained devices. For simple sensors, application-layer monitoring may be unnecessary because no such traffic is generated:
    \begin{equation}
    \sum_{l=1}^{\beta_i} y_{i,l} = 1, \quad \forall i \in F
    \end{equation}
\end{enumerate}

\section{Experimental Evaluation}

\begin{figure}
    \centering
    \includegraphics[width=0.8\textwidth]{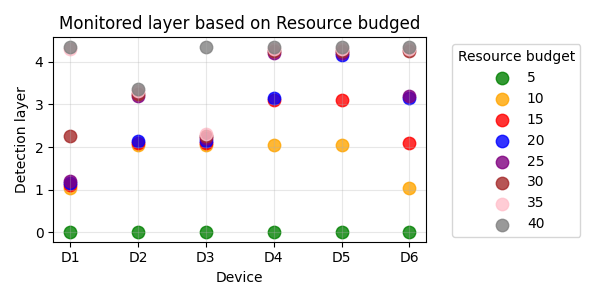}
    \caption{Device layer monitoring based on resource allocation.}
    \label{fig:fig1}
\end{figure}

\begin{figure}
    \centering
    \includegraphics[width=0.8\textwidth]{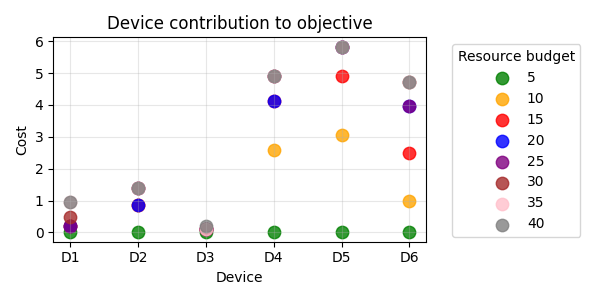}
    \caption{Device contribution to objective based on resource budget.}
    \label{fig:fig2}
\end{figure}

Solving Constraint Integer Programs (SCIP) is a state-of-the-art open-source optimization framework designed for solving
mixed-integer programming (MIP) and mixed-integer nonlinear programming (MINLP) problems. Developed primarily at the Zuse
Institute Berlin, SCIP integrates techniques from constraint programming, integer programming, and SAT solving into a
unified branch-and-bound framework. It is particularly known for its flexibility, allowing users to customize components
such as branching rules, cutting planes, and heuristics, making it suitable for both academic research and industrial
applications. SCIP supports a wide range of problem formulations and is often used as a benchmark solver due to its strong
performance and extensibility \cite{scip2009solving}.

The model is instantiated on a small heterogeneous network in order to illustrate its behavior. The four monitoring
layers $l \in \{1, 2, 3, 4\}$ are associated with detection rates $d_1 = 0.2$, $d_2 = 0.5$, $d_3 = 0.8$, $d_4 = 0.95$
and with monitoring costs $c_1 = 1$, $c_2 = 2$, $c_3 = 4$, $c_4 = 7$, reflecting the assumption that deeper inspection
yields higher detection but at a higher resource cost. The total monitoring budget is set to $R \in \{5, 10, 15, 20, 25, 30, 35, 40\}$. Critical devices
must be monitored at least up to layer $\alpha = 2$, with the critical set defined as $C = \{\textit{dev\_2},
\textit{dev\_3}, \textit{dev\_4}\}$, while feasibility limits restrict $\textit{dev\_2}$ to a maximum monitoring depth
of $\beta_{\textit{dev\_2}} = 3$. The network contains six devices with importance weights $w_i$ and attack
probabilities $p_i$ as follows: $\textit{dev\_1}$ ($w = 1$, $p = 0.998$), $\textit{dev\_2}$ ($w = 3$, $p = 0.579$),
$\textit{dev\_3}$ ($w = 5$, $p = 0.045$), $\textit{dev\_4}$ ($w = 10$, $p = 0.517$), $\textit{dev\_5}$ ($w = 9$,
$p = 0.682$), and $\textit{dev\_6}$ ($w = 7$, $p = 0.71$). This configuration is used as input to the SCIP solver to
evaluate the proposed allocation model.

In Fig. \ref{fig:fig1}, the optimal monitoring layer allocation for each device is shown, while in Fig.
\ref{fig:fig2}, the contribution of each device to the objective based on the resource budget is illustrated, highlighting how critical devices
(dev\_2, dev\_3, dev\_4) are monitored at least up to layer 2, while dev\_2 is limited to a maximum of layer 3. The allocation
balances detection efficiency and resource usage across the heterogeneous network. For a small resource budget,
the model prioritizes monitoring the most important and high-risk devices at deeper layers, while for a larger budget,
it allows for broader monitoring across more devices and layers. If resource budget is too small, the model yields
no solution (R=5).

\section{Conclusion}
This paper introduced a resource-aware allocation model for layered intrusion detection in heterogeneous
networks, formulated as an integer linear program in which each device is assigned a single monitoring depth
ranging from Ethernet to the application layer. The model jointly captures device importance, attack
probability, layer-dependent detection rates, per-layer monitoring costs, a global resource budget, minimum
monitoring requirements for critical devices, and maximum-feasibility limits for constrained devices.

The formulation was solved with the SCIP optimization framework on an illustrative six-device network. The
resulting allocation respected the budget and feasibility constraints, ensured that critical devices were
monitored at least up to the required layer, and concentrated deeper inspection on devices with higher
importance and attack probability, while limiting monitoring overhead on low-value or constrained devices.
Overall, the approach offers a structured and reproducible way to balance detection efficiency and resource
usage across heterogeneous networks. Future work will consider larger and more realistic network instances,
uncertainty in attack probabilities, and the integration of this allocation layer with the machine-learning-based
detectors discussed in the related work.

\bibliographystyle{ieeetr}
\bibliography{Bibliografia}
\end{document}